\documentclass{PoS}

\title{Charm quark system on the physical point in $2+1$ flavor lattice QCD}

\ShortTitle{Charm quark system on the physical point in $2+1$ flavor lattice QCD}

\author{
 \speaker{Y. Namekawa} \thanks{E-mail: namekawa@ccs.tsukuba.ac.jp}
 for PACS-CS Collaboration,
 \\
 \llap{}Center for Computational Sciences, University of Tsukuba, Tsukuba,
 Ibaraki 305-8577, Japan
}

\abstract{
We investigate the charm quark system
on 2+1 flavor PACS-CS configurations.
Calculations are performed at the lattice spacing
$a^{-1}=2.194(10)$ GeV and the spatial extent $L=2.9$ fm
with O(a)-improved Wilson fermions for the light quarks
and the relativistic heavy fermion for the charm quark.
Our dynamical $ud$ and strange quark masses and valence
charm quark mass are set to their physical values.
A comparison of the mass spectrum and decay constants
with experiments is presented.
Our results for the charm quark mass and CKM matrix elements
are also reported.
          }
\FullConference{ The XXIX International Symposium on Lattice Field Theory - Lattice 2011\\
July 10-16, 2011\\
Squaw Valley, Lake Tahoe, California}

\begin{document}

\section{Introduction}
\label{section:introduction}

Precise determination of the Cabibbo-Kobayashi-Maskawa (CKM)
quark mixing matrix is an indispensable step to establish
the validity range of the standard model,
and to search for new physics at higher energy scales.
Lattice QCD has been making steady progress in this direction.
%

One of the difficulties with the charm quark in lattice QCD simulations
at a typical cutoff $a^{-1} \approx 2$~GeV 
resides in significant cutoff errors due to the charm quark mass.
The heavy quark mass correction is $m_Q a \sim O(1)$,
and hence it is desirable to control $m_Q a$ errors
to achieve an accuracy of a few percent.
The Fermilab action~\cite{Fermilab_action} and
the relativistic heavy quark action~\cite{RHQ_action_Tsukuba,RHQ_action_Columbia}
have been proposed to meet this goal.
The relativistic heavy quark action
removes the leading cutoff errors of $O((m_Q a)^n)$
and the next to leading effects of $O((m_Q a)^n (a \Lambda_{QCD}))$
with arbitrary order $n$,
once all of the parameters in the heavy quark action
are determined nonperturbatively.
%

Another source that prevents precise evaluations in lattice QCD 
is the error associated with chiral extrapolations
in the light quark masses.
This problem has been increasingly alleviated through progress
toward simulations with lighter and lighter dynamical quark masses.
%
The acceleration of dynamical lattice QCD simulation
using multi-time steps
for infrared and ultraviolet modes 
has made it possible
to run simulations with light up, down, and strange quark masses
around their physical values~\cite{PACS_CS_0}.   
%
%
%
In fact, we can proceed one more step and reweight 
dynamical simulations
such that dynamical quark masses take exactly 
the physical values~\cite{PACS_CS_1}.
%
%
Once the reweighting is successfully made,
ambiguities associated with chiral extrapolations are
completely removed.

In this paper, we present our work for the charm quark system
treated with the relativistic heavy quark formalism
on the 2+1 dynamical flavor PACS-CS configurations
of $32^3 \times 64$ lattice generated 
with the Wilson-clover quark
and reweighted to the physical point for 
up, down and strange quark masses. 
%
%
%
Details of our calculations are presented in Ref.~\cite{PACS_CS_2011}.

\section{Set up}
\label{section:setup}

Our calculation is based on a set of $N_f=2+1$ flavor
dynamical lattice QCD configurations
generated by the PACS-CS Collaboration~\cite{PACS_CS_1}
on a $32^3\times 64$ lattice 
using the nonperturbatively $O(a)$-improved Wilson quark action 
with $c_{\rm SW}^{\rm NP}=1.715$~\cite{Csw_NP}
and the Iwasaki gauge action 
at $\beta=1.90$.
The aggregate of 2000 MD time units were generated at the hopping parameter
given by $(\kappa_{ud}^0,\kappa_{s}^0)=(0.13778500, 0.13660000)$,
and 80 configurations at every 25 MD time units were used for measurements.
We then reweight those configurations to the physical point
given by $(\kappa_{ud},\kappa_{s})=(0.13779625, 0.13663375)$.
The reweighting shifts the masses of $\pi$ and $K$ mesons 
from $m_\pi=152(6)$~MeV and $m_K=509(2)$~MeV
  to $m_\pi=135(6)$~MeV and $m_K=498(2)$~MeV,
with the cutoff at the physical point estimated
to be $a^{-1}=2.194(10)$~GeV.


Our relativistic heavy quark action~\cite{RHQ_action_Tsukuba} is given by
\begin{eqnarray}
 S_Q
 &=& \sum_{x,y}\overline{Q}_x D_{x,y} Q_y,\\
 D_{x,y}
 &=& \delta_{xy}
     - \kappa_{Q}
       \sum_i \left[  (r_s - \nu \gamma_i)U_{x,i} \delta_{x+\hat{i},y}
                     +(r_s + \nu \gamma_i)U_{x,i}^{\dag} \delta_{x,y+\hat{i}}
              \right]
     \nonumber \\
 &&  - \kappa_{Q}
              \left[  (r_t -     \gamma_4)U_{x,4} \delta_{x+\hat{4},y}
                     +(r_t +     \gamma_4)U_{x,4}^{\dag} \delta_{x,y+\hat{4}}
              \right]
     \nonumber \\
 &&  - \kappa_{Q}
              \left[   c_B \sum_{i,j} F_{ij}(x) \sigma_{ij}
                     + c_E \sum_i     F_{i4}(x) \sigma_{i4}
              \right] \delta_{xy},
\end{eqnarray}
where $\kappa_Q$ is the hopping parameter for the heavy quark.
The parameters $r_t, r_s, c_B, c_E$, and $\nu$ are adjusted as follows.
We are allowed to choose $r_t=1$,
and we employ a one-loop perturbative value for $r_s$~\cite{RHQ_parameters}.
For the clover coefficients $c_B$ and $c_E$,
we include the non-perturbative contribution
in the massless limit $c_{\rm SW}^{\rm NP}$
for three flavor dynamical QCD~\cite{Csw_NP},
and calculate the heavy quark mass dependent contribution 
to one-loop order in perturbation theory~\cite{RHQ_parameters}.
%
The parameter $\nu$ is determined non-perturbatively
to reproduce the relativistic dispersion relation for
the spin-averaged  $1S$ states of the charmonium.
Writing 
\begin{equation}
   E({\vec p})^2
 = E({\vec 0})^2+c_{\rm eff}^2 |{\vec p}|^2,
\end{equation}
for $|{\vec p}| = 0, (2 \pi / L), \sqrt{2} (2 \pi / L)$,
and demanding the effective speed of light $c_{\rm eff}$
to be unity, we find $\nu=1.1450511$ with which
we have $c_{\rm eff} = 1.002(4)$.
%
%
It is noted that the cutoff error of
$O(\alpha_s^2 (a \Lambda_{QCD}))$
remains,
in addition to
$O((a \Lambda_{QCD})^2)$,
due to the use of one-loop perturbative values in part
for the parameters of our heavy quark action.

We tune the heavy quark hopping parameter to reproduce  
an experimental value of the mass for the spin-averaged $1S$ states of the charmonium,
given by
%
$M(1S)^{exp}
 = (M_{\eta_c} + 3 M_{J/\psi})/4
 = 3.0678(3) \mbox{~GeV~\cite{PDG_2010}}.
$
%
This leads to $\kappa_{\rm charm}=0.10959947$
for which our lattice QCD measurement
yields the value $M(1S)^{lat} = 3.067(1)(14)$~GeV,
where the first error is statistical,
and the second is a systematic
from the scale determination.
Our parameters for the relativistic heavy quark action
are summarized in Table~\ref{table:input_parameters_for_RHQ}.

Statistical errors are analyzed by the jackknife method
with a bin size of 100 MD time units (4 configurations),
as in the light quark sector~\cite{PACS_CS_1}.

\begin{table}[t]
\caption{
 Simulation parameters.
 MD time is the number of trajectories
 multiplied by the trajectory length.
}
\label{table:statistics}
\begin{center}
\begin{tabular}{ccccc}
\hline
 $\beta$           &
 $\kappa_{\rm ud}$ & $\kappa_{\rm s}$ &
 \# conf           & MD time
\\ \hline
 1.90              &
 0.13779625        & 0.13663375 &
 80                & 2000
\\ \hline
\end{tabular}
\end{center}
\end{table}

\begin{table}[t]
\caption{
 Parameters for the relativistic heavy quark action.
}
\label{table:input_parameters_for_RHQ}
\begin{center}
\begin{tabular}{cccccc}
\hline
 $\kappa_{\rm charm}$  & $\nu$     & $r_s$     & $c_B$     & $c_E$
\\ \hline
 0.10959947            & 1.1450511 & 1.1881607 & 1.9849139 & 1.7819512
\\ \hline
\end{tabular}
\end{center}
\end{table}


\section{Charmonium spectrum and charm quark mass}
\label{section:result_1}

\begin{figure}[t]
\begin{center}
 \includegraphics[width=75mm]{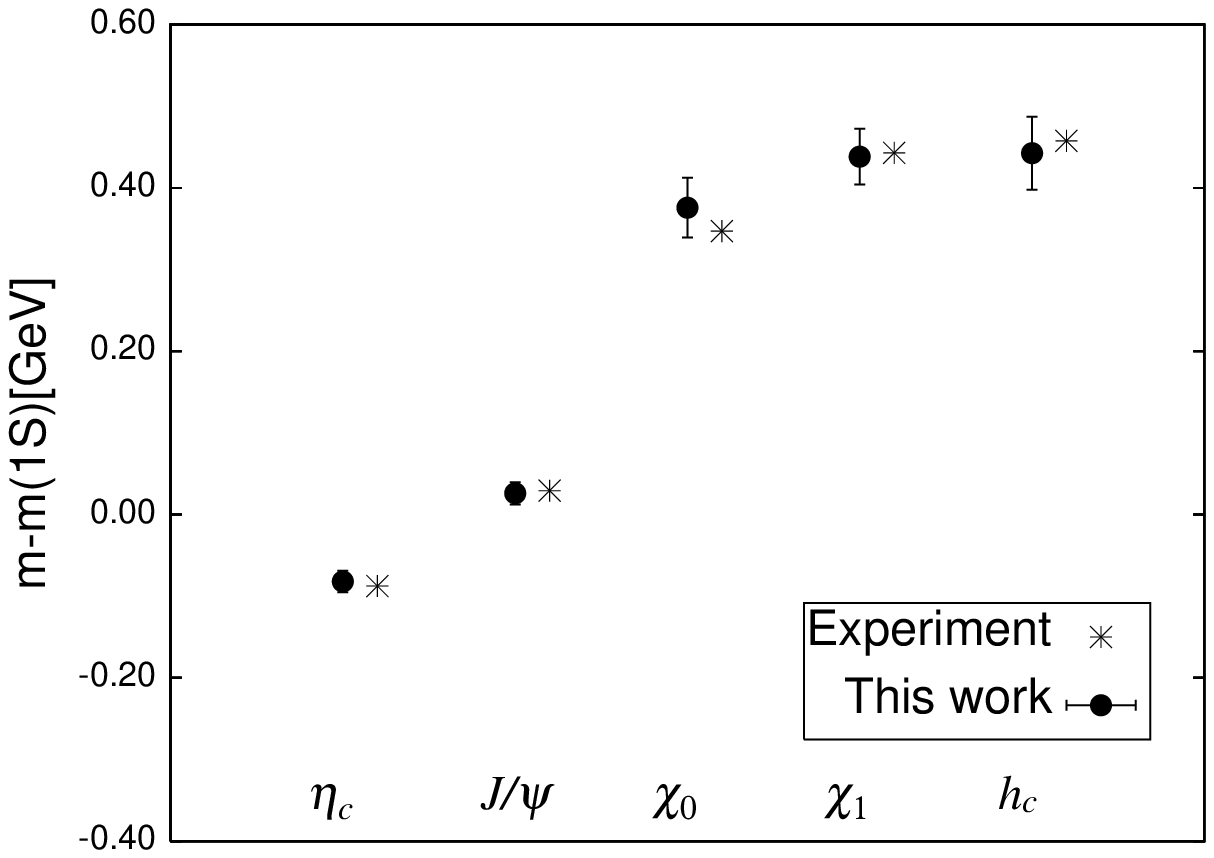}
 \includegraphics[width=75mm]{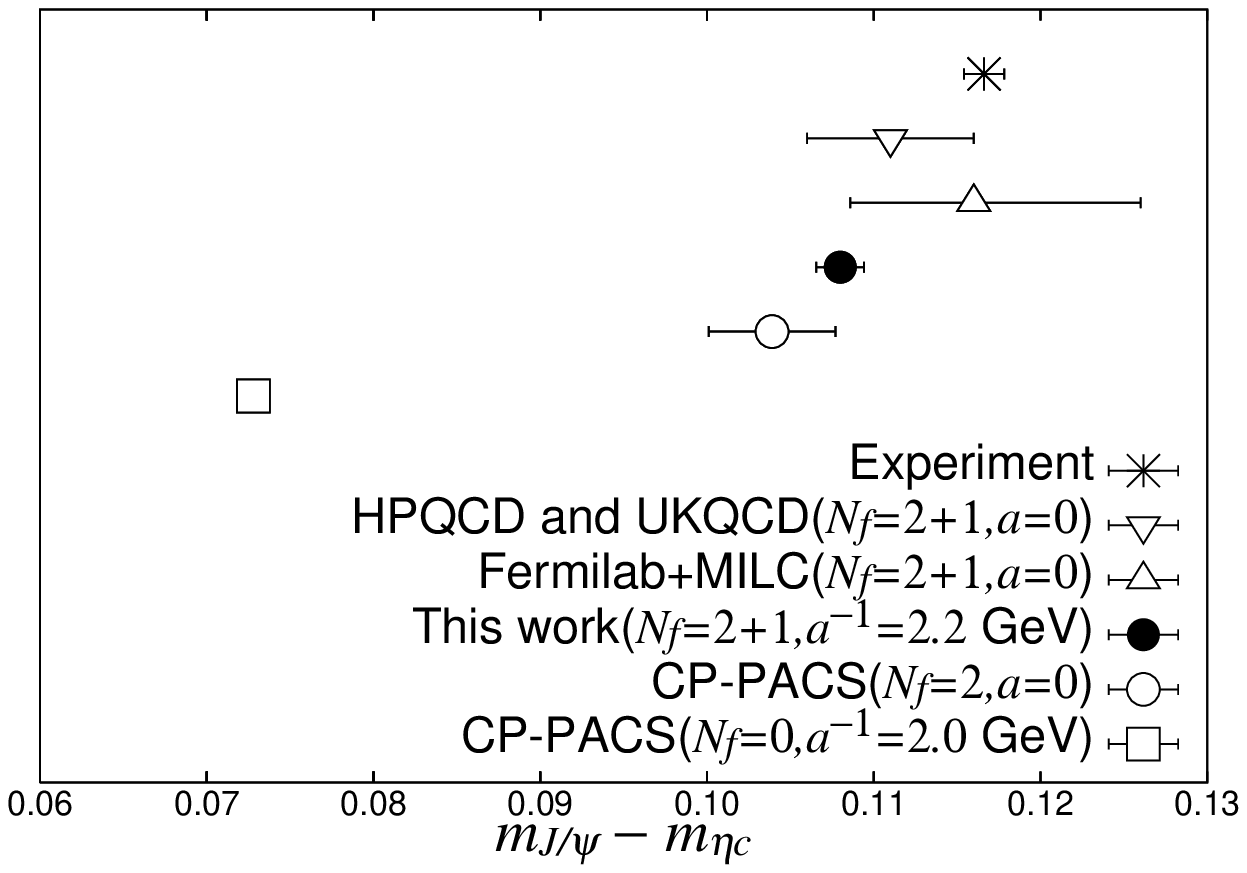}
 \caption{
  Our results for the charmonium mass spectrum(left panel)
  and the hyperfine splitting of the charmonium
  with different number of flavors(right panel).
 }
 \label{figure:mass_charmonium_all}
\end{center}
\end{figure}

Our results for the charmonium spectrum
on the physical point are summarized in
Fig.~\ref{figure:mass_charmonium_all}.
%
Within the error of 0.5--1\%,
the predicted spectrum is in reasonable agreement with experiment.

Let us consider the $1S$ states more closely.
Since these states are employed to tune the charm quark mass, 
the central issue here is the magnitude of the hyperfine splitting.
Our result $m_{J/\psi}-m_{\eta_c}=0.108(1)(0)$~GeV,
where the first error is statistical
and the second error is systematic from the scale determination,
is 7\% smaller than the experimental value of 0.117~GeV.  
%
In Fig.~\ref{figure:mass_charmonium_all},
we compare the present result
on $N_f=2+1$ flavor dynamical configurations 
with previous attempts on $N_f=2$ dynamical
and quenched configurations
using the same heavy quark formalism
and the Iwasaki gluon action~\cite{RHQ-N_f_0_2}.
Other results by recent lattice QCD
simulations by Fermilab lattice and MILC Collaborations~\cite{FNAL_2010},
HPQCD and UKQCD Collaborations~\cite{HPQCD_2007_2010}
are also plotted.
We observe a clear trend that
incorporation of dynamical light quark effects 
improves the agreement.

We should note that we have not evaluated our systematic errors
for the hyperfine splitting, yet.
The continuum extrapolation needs to be performed.
A naive order counting implies that the cutoff effects of
$O(\alpha_s^2 (a \Lambda_{QCD}), (a \Lambda_{QCD})^2)$
from the relativistic heavy quark action
are at a percent level.
Another aspect is that
dynamical charm quark effects
and disconnected loop contributions,
albeit reported to give a shift of only a few MeV~\cite{disconnected},
are not included in the present work.
Additional calculations are needed to draw a definite conclusion
for the hyperfine splitting of the charmonium spectrum.
We leave it for a future work.

\begin{figure}[t]
\begin{center}
 \includegraphics[width=75mm]{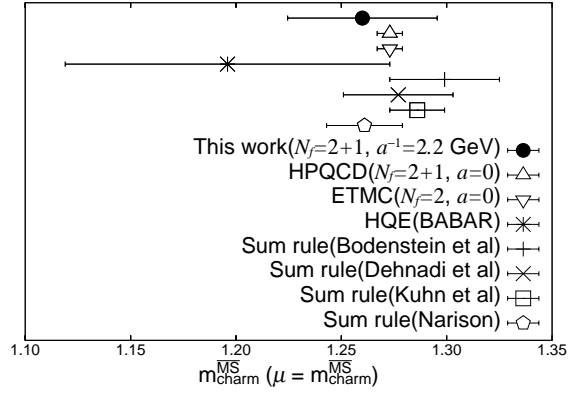}
 \caption{
 Comparison of the charm quark mass.
 We employ $N_f=4$ running in this plot.
 }
 \label{figure:m_charm}
\end{center}
\end{figure}

Using the axial vector Ward-Takahashi identity,
the charm quark mass is obtained.
%
The systematic error due to the heavy quark of
$O(\alpha_s^2 (a \Lambda_{QCD}), (a \Lambda_{QCD})^2)$
will be estimated by data on finer lattices in the future.
Figure~\ref{figure:m_charm} compares our result
with a recent $N_f=2+1$ lattice QCD estimation
by the HPQCD Collaboration~\cite{HPQCD_2007_2010}.
%
Another result by ETM Collaboration
is also plotted~\cite{ETMC_2009_2011}.
In addition to lattice QCD determinations,
recent continuum results
using the Heavy Quark Expansions(HQE)~\cite{BaBar},
as well as sum rules
~\cite{Sum_rule_1},
are shown.
All these results are consistent.

\section{Charmed meson and charmed-strange meson spectrum}
\label{section:result_2}

We calculate the charmed meson
and charmed-strange meson masses
which are stable on our lattice with the spatial size of $L = 2.88(1)$~fm
and a lattice cutoff of $a^{-1}=2.194(10)$~GeV. 
The $D^*$ and $D_s^*$ meson decay channels are not open
in our lattice setup.
$D_{s0}^*$ and $D_{s1}$ meson masses are
below the $D K$ threshold~\cite{PDG_2010} but above 
the $D_s \pi$ threshold.
Their decays, however, 
are prohibited by the isospin symmetry.
On the other hand,
$D_{0}^*$ and $D_1$ meson masses are not computed
since their decay channels are open,
and therefore a calculation involving $D \pi$ contributions is needed.

\begin{figure}[t]
\begin{center}
 \includegraphics[width=75mm]{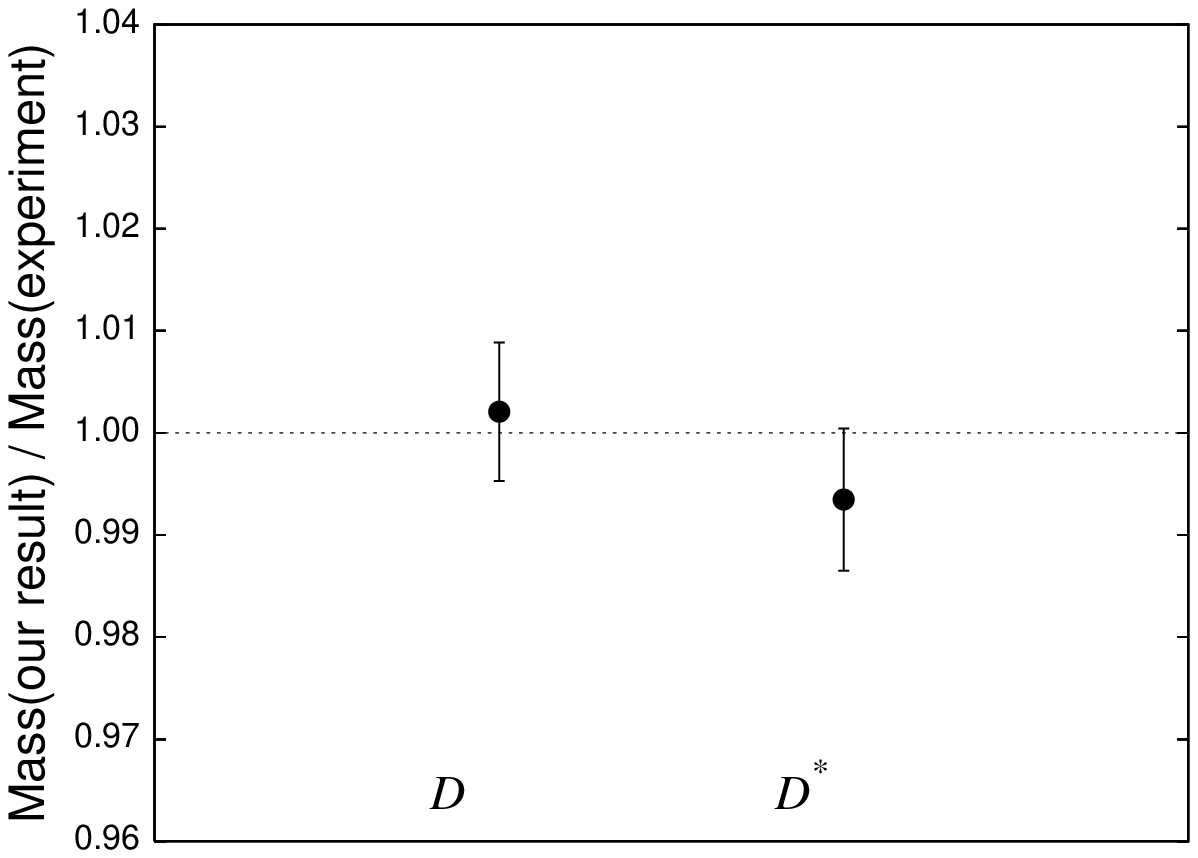}
 \includegraphics[width=75mm]{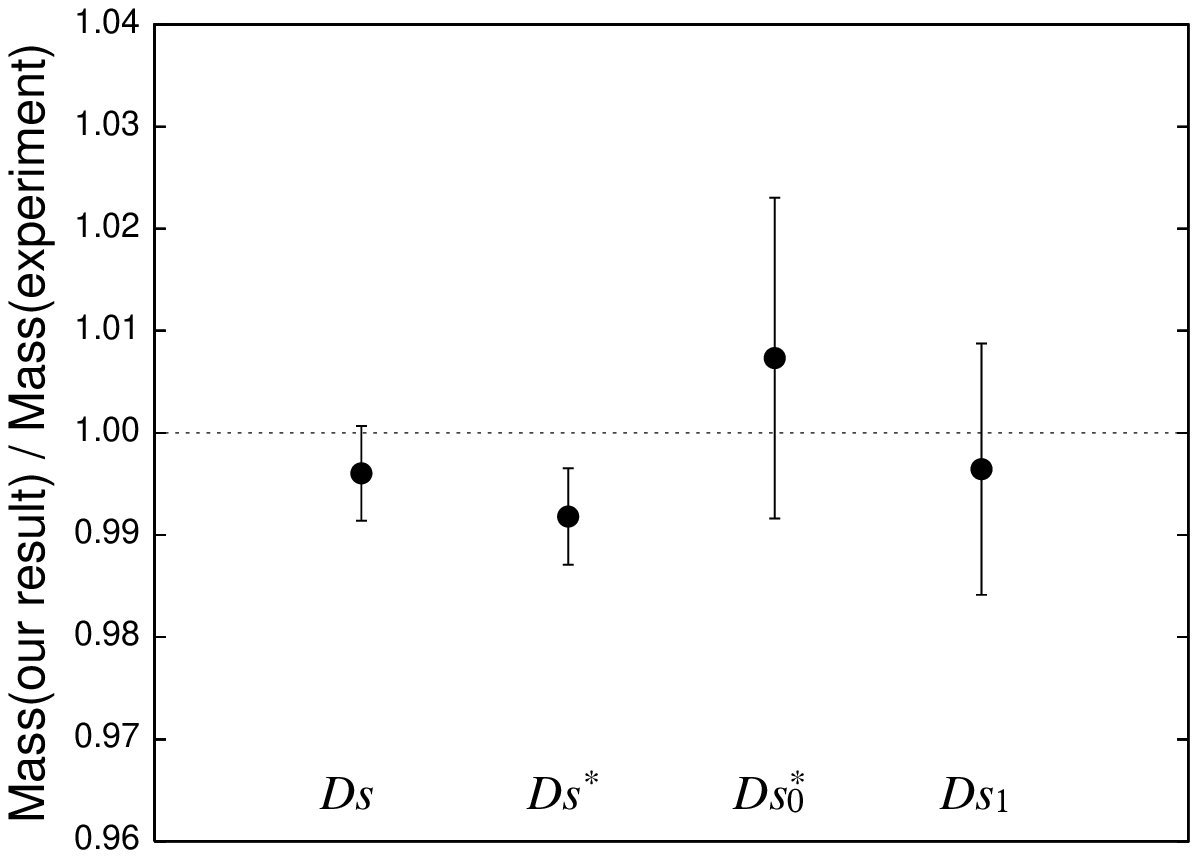}
 \caption{
 Our results for charmed meson masses(left panel)
 and charmed-strange meson masses(right panel)
 normalized by the experimental values.
 }
 \label{figure:mass_ud_charm}
\end{center}
\end{figure}

Our results are summarized in Fig.~\ref{figure:mass_ud_charm}.
%
All our values for the heavy-light meson quantities are predictions,
because the physical charm quark mass has already been fixed
with the charmonium spectrum.
The experimental spectrum are reproduced in $2 \sigma$ level.
The potential model predicts
the $D_{s0}^*$ meson mass is
above the $D K$ threshold~\cite{potential_model},
which deviates from the experiment significantly~\cite{D_s0_review}.
%
%
Our result, however, does not indicate such a large difference
from the experimental value.
A similar result is obtained in other lattice QCD calculations~\cite{chi_QCD,TRIUMF}.
The $D_{s0}^*$ meson mass is below the $D K$ threshold.
%
%
It should be noticed that our calculation does not cover
$D K$ scattering states yet.
A $D K$ contamination for $D_{s0}^*$ and $D_{s1}$ meson masses
could be considerably large.
Further analysis is required to validate
our results for $D_{s0}^*$ and $D_{s1}$ meson spectrum.

%
%
%

%
%
%

\begin{figure}[t]
\begin{center}
 \includegraphics[width=75mm]{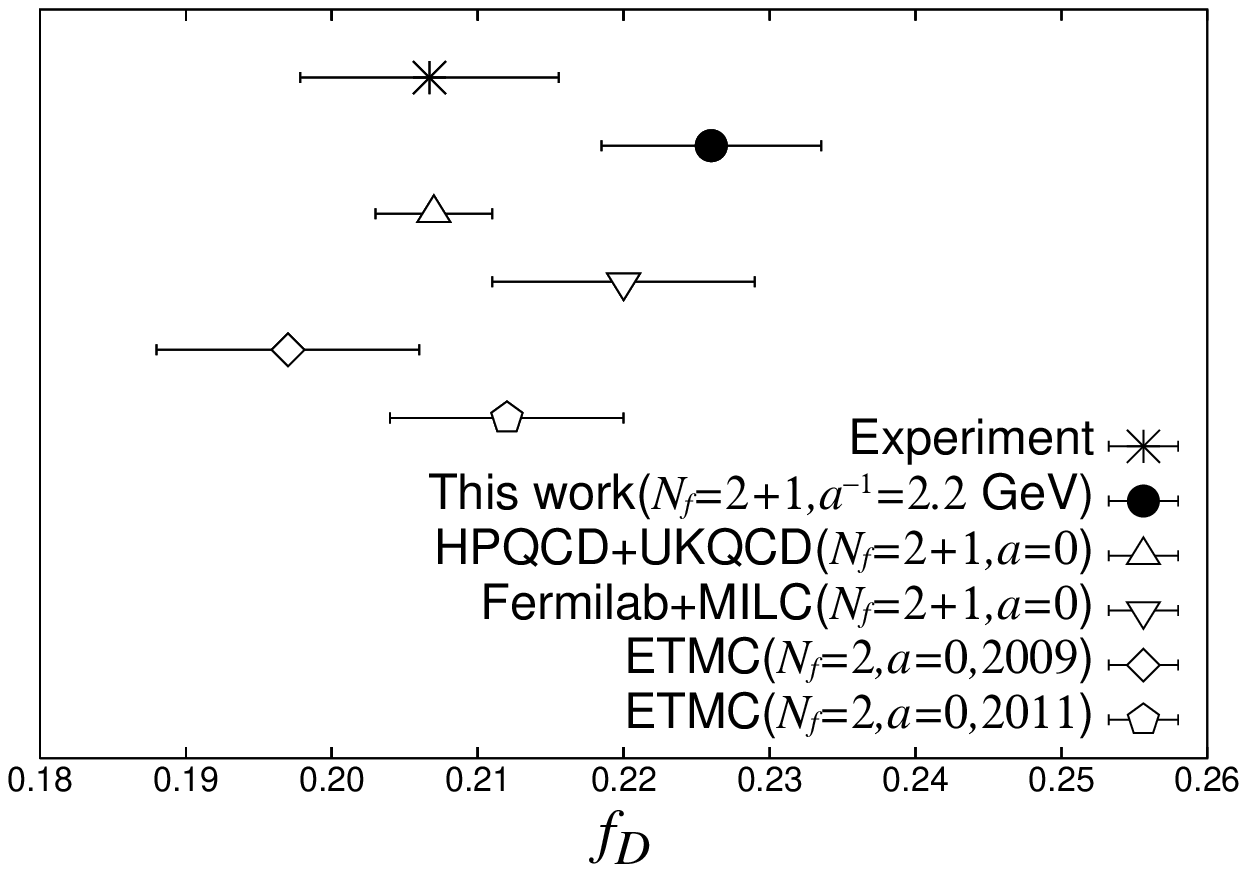}
 \includegraphics[width=75mm]{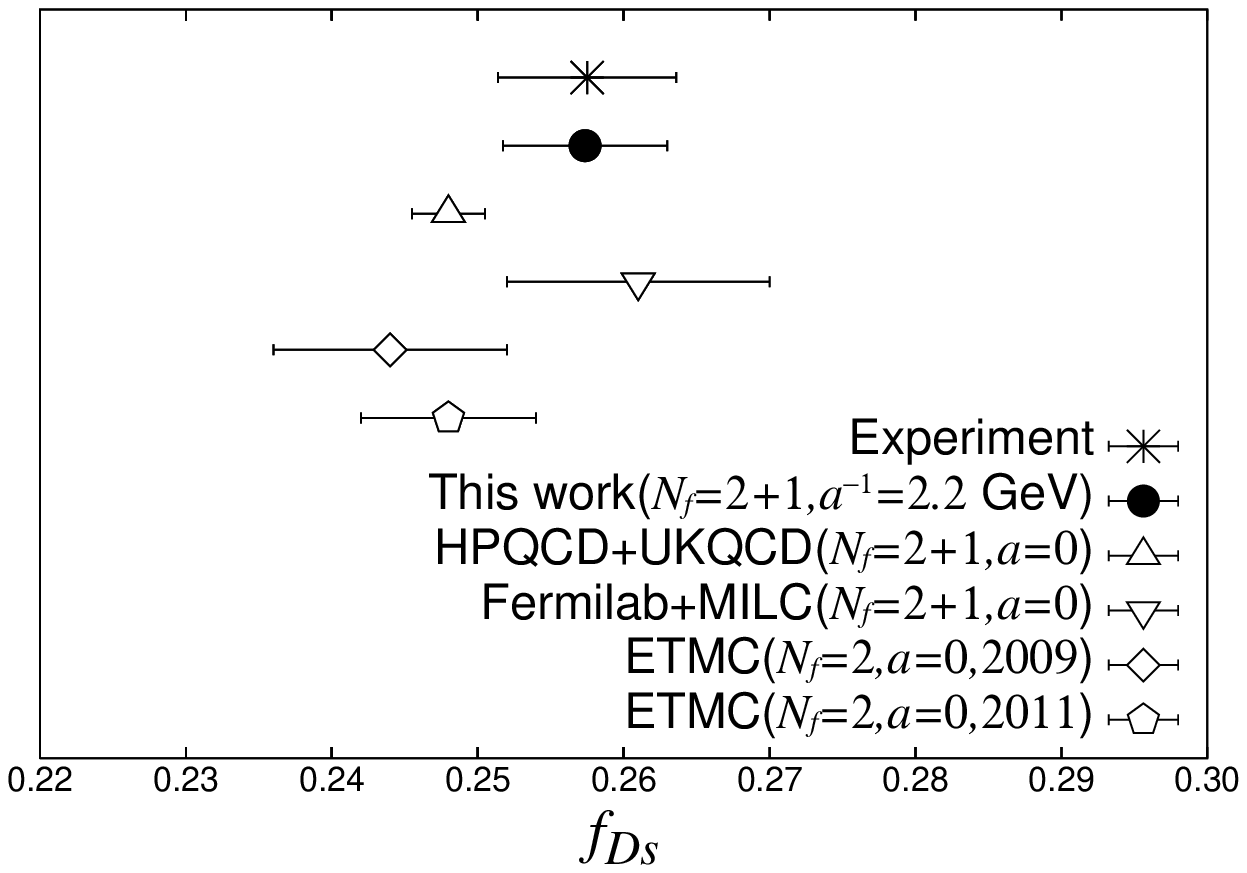}
 \caption{
 Comparison of pseudoscalar decay constants for
 the charmed meson (left panel)
 and charmed-strange meson (right panel).
 }
 \label{figure:f_PS_ud_charm_and_f_PS_s_charm}
\end{center}
\end{figure}

%
%
%
%
Figure~\ref{figure:f_PS_ud_charm_and_f_PS_s_charm}
shows our decay constants including the experimental values~\cite{PDG_2010},
as well as three recent lattice QCD results: 
HPQCD and UKQCD Collaboration~\cite{HPQCD_2007_2010},
Fermilab lattice and MILC group~\cite{FNAL_2010}, and
ETM Collaboration~\cite{ETMC_2009_2011}.
%
Our value for $f_{D_s}$ is in accordance
with experiment,
while that for $f_D$ is somewhat larger. 
Comparing four sets of lattice determinations,
we observe, both for $f_D$ and $f_{D_s}$,  
an agreement between our values and those of the Fermilab group,
while there seems to be a discrepancy between our values and
those by the HPQCD and UKQCD Collaboration and ETM Collaboration,
though continuum extrapolation is needed on our part. 
%



The standard model relates $|V_{cd}|$
to the leptonic decay width of the $D$ meson 
$\Gamma(D \rightarrow l \nu)$ by
\begin{eqnarray}
 \Gamma(D \rightarrow l \nu)
 = \frac{G_F^2}{8 \pi} f_{D}^2 m_l^2 m_{D}
   \left( 1 - \frac{m_l^2}{m_{D}^2} \right)^2 |V_{cd}|^2,
\end{eqnarray}
where $G_F$ is the Fermi coupling constant,
and $m_l$ is the lepton mass in the final state.
A lattice determination of the $D$ meson decay constant $f_{D}$
with the experimental value of $\Gamma(D \rightarrow l \nu)$
gives $|V_{cd}|$.
$|V_{cs}|$ can be obtained in the same way.

\begin{figure}[t]
\begin{center}
 \includegraphics[width=75mm]{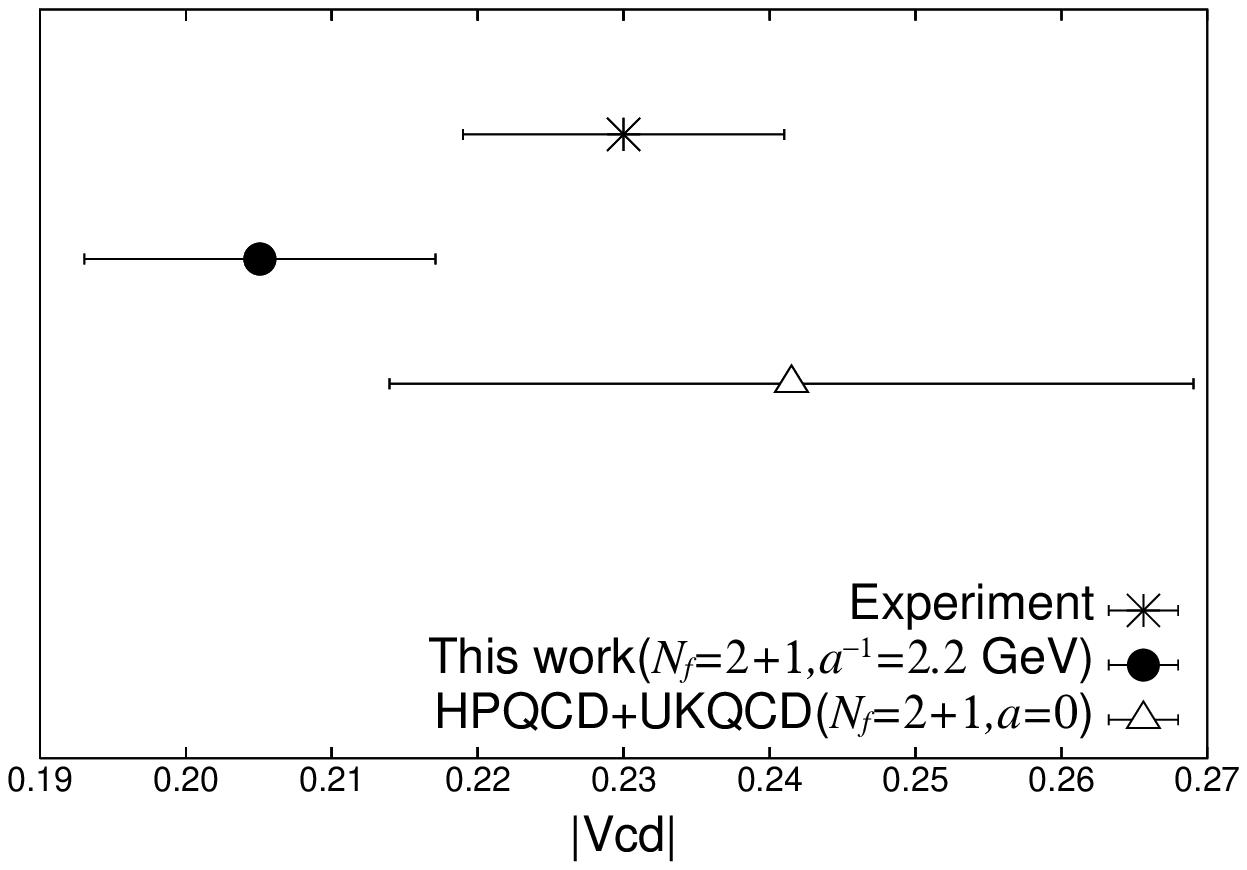}
 \includegraphics[width=75mm]{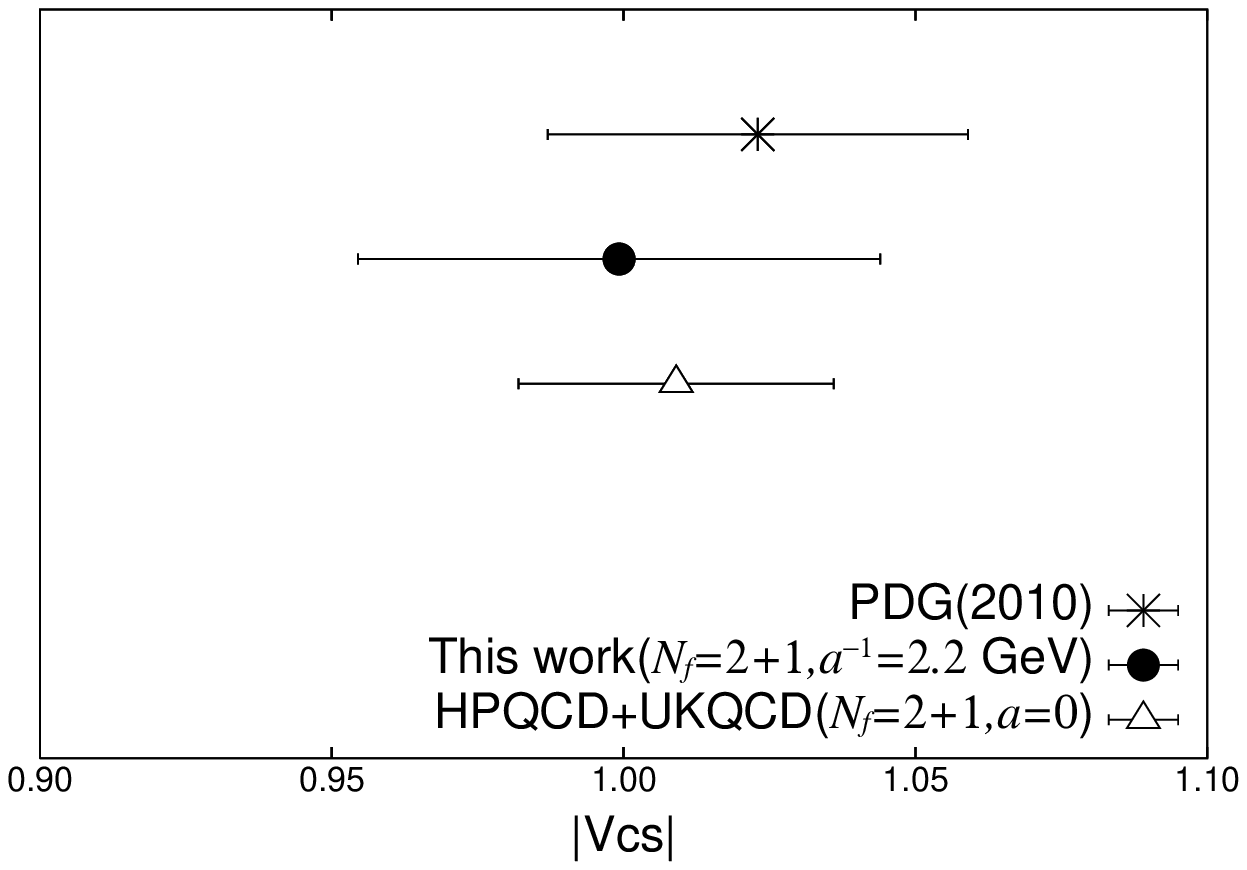}
 \caption{
 Comparison of the CKM matrix elements,
 $|V_{cd}|$ (left panel) and $|V_{cs}|$ (right panel).
 }
 \label{figure:V_cd_and_V_cs}
\end{center}
\end{figure}

We estimate $|V_{cd}|$ from our $D$ meson mass and decay constant
with the CLEO value of $\Gamma(D \rightarrow l \nu)$~\cite{CLEO_1}.
Up to our heavy quark discretization error of
$O(\alpha_s^2 (a \Lambda_{QCD}), (a \Lambda_{QCD})^2)$, 
we obtain
$|V_{cd}|$.
%
Our result of $|V_{cd}|$
is about 10\% smaller than the PDG value~\cite{PDG_2010},
as shown in Fig.~\ref{figure:V_cd_and_V_cs}.
Similarly, using the CLEO value of
$\Gamma(D_s \rightarrow l \nu)$
~\cite{CLEO_1},
we find $|V_{cs}|$,
which is consistent with PDG.
%



\begin{acknowledgments}
Y.N. thanks Takayuki Matsuki for helpful comments.
Numerical calculations for the present work have been carried out
on the PACS-CS computer
under the ``Interdisciplinary Computational Science Program'' of
the Center for Computational Sciences, University of Tsukuba.
This work is supported in part by Grants-in-Aid of the Ministry
of Education, Culture, Sports, Science and Technology-Japan
 (Nos. 18104005, 20340047,
 20540248, 21340049, 22244018, and 22740138),
and Grant-in-Aid for Scientific Research on Innovative Areas
(No. 2004: 20105001, 20105002, 20105003, 20105005, and
 No. 2104: 22105501)
\end{acknowledgments}


\end{document}